# Mapping open quantum dynamics onto graphs

Kyuho Kim[1], Dayeong Lee[1], Seungkyun Park[1,2,4], Xianji Piao[3§], Namkyoo Park[2†], and Sunkyu Yu[1*]

[1]Intelligent Wave Systems Laboratory, Department of Electrical and Computer Engineering, Seoul National University, Seoul 08826, Korea

[2]Photonic Systems Laboratory, Department of Electrical and Computer Engineering, Seoul National University, Seoul 08826, Korea

[3]Wave Engineering Laboratory, School of Electrical and Computer Engineering, University of Seoul, Seoul 02504, Korea

[4]KAIST InnoCORE PICORE Center, Korea Advanced Institute of Science and Technology, Daejeon, 34141, Korea

E-mail address for correspondence: [§]piao@uos.ac.kr, [†]nkpark@snu.ac.kr, [*]sunkyu.yu@snu.ac.kr

**Abstract**

Graph-theoretic frameworks have been widely employed in quantum physics to address the high-dimensional complexity of quantum systems. Although open quantum dynamics incorporates system-bath coupling via numerous interacting operators, it has been formulated algebraically with a partial set of jump operators or statistically universal reservoirs, leaving the underlying connectivity structure largely unexplored. Here, we propose a universal graph-theoretic framework for Markovian quantum dynamics. The framework maps open quantum dynamics onto two uniquely defined graphs, where the quantum master equation is rigorously interpreted as the average wave characteristic of operator-valued signals across the graphs. Applying this framework to the open quantum Rabi model, we demonstrate an open-system generalization of Fock-state lattices, characterize graph-topological signatures of dissipation, and classify the weak-to-ultrastrong coupling transition. Building on these representations, graph pruning reveals the backbone of open quantum dynamics, which enables superior graph neural-network learning. Our results bridge graph theory and open quantum dynamics, achieving efficient data-driven analysis of high-dimensional complexity.

## Introduction

Graph-theoretic frameworks—representing complex interactions through vertices and edges—offer a powerful language for abstracting complex physical phenomena, serving to clarify emergent collective dynamics[1,2], and to realize efficient design through graph-based learning[3-6]. Quantum systems are intrinsically suited to this perspective, because many of their behaviours are governed by diverse collective interactions, ranging from coherent couplings and transitions to correlations and entanglement. Accordingly, graph-theoretic frameworks have been widely employed in quantum physics, especially in high-dimensional systems, such as Fock-state lattices in quantum electrodynamics[7-9], quantum graphs in scattering phenomena[10], and tensor networks in many-body systems[11].

Open quantum dynamics is likewise naturally suited to graph representation. As described by the Lindblad quantum master equation (QME)[12,13], an $N$-dimensional open quantum system is represented in an operator space whose complexity scales as $N^2$. Furthermore, the system-bath coupling is encoded in the intricate bath-induced mixing of these operators once the environment is traced out. Despite the necessity of abstracting such high-dimensional complexity, the dynamics of open quantum systems has been treated algebraically, with most studies focusing on only a small number of jump operators[13,14]. Although some works have examined the statistical distributions of the bath-induced mixing of coupling operators—ranging from normal distributions[15-20] to exotic forms beyond the central limit theorem[21-23]—the underlying connectivity structure of system-bath coupling remains largely unexplored. In this context, a natural question arises: is there a universal graph-theoretic framework that can consistently abstract both isolated and open quantum dynamics?

Here, we propose the Schrödinger operator graph—a universal graph framework applicable to both isolated and open quantum systems. The framework is built on operator-valued edge signals on a magnetic graph[24,25]—a graph with conjugate-symmetric phase factors assigned to its edges—together with vertex potentials. We reveal that the graph-space averages of signal wave features yield representations mathematically equivalent to Markovian quantum dynamics. We apply this framework to the open quantum Rabi model (QRM), revealing distinct graph signatures of relaxation and dephasing mechanisms as well as a graph-based classification of the transition from the weak-coupling to the ultrastrong-coupling (USC) regime. We also develop a pruning method for graph sparsification that provides physically allowed graph approximations. Building on the critical roles of hub edges and vertex potentials revealed by the pruning, we devise a graph convolutional network (GCN) for inferring the Liouville spectra of random open QRMs from their abstracted graphs. The result opens new avenues for graph-based modelling of high-dimensional quantum physics.

## Results

### Schrödinger operator graphs

As a mathematical framework for mapping quantum dynamics onto graphs, we establish the Schrödinger operator graph, which is designed to analyse wave characteristics of physically valid, operator-valued signals across the graph. Consider an $N_G$-vertex, fully connected, and weighted graph, $G = (V, E)$, where $V$ and $E$ denote the sets of vertices and edges, respectively (Fig. 1a). To reflect the reciprocity inherent in physical systems, we restrict $G$ to the class of magnetic graphs [24,26], characterized by an $N_G \times N_G$ complex-valued adjacency matrix $A(G)$ satisfying $[A(G)]_{pq} = [A(G)]_{qp}^*$ and $[A(G)]_{pp} = 0$. Motivated by the Schrödinger equation, we define the graph

Hamiltonian $H(G) = \Delta(G) + \Phi(G)$—which is an element of an $N_G^2$-dimensional Hilbert-Schmidt operator space $\mathcal{H}_G$—where $\Delta(G)$ is the magnetic Laplacian of $G$ (see Methods)[24,26] and $\Phi(G)$ is a diagonal matrix representing the vertex potential. The Hamiltonian operator $H(G)$ governs the wave features of a signal across $G$, by reflecting the graph topology via $\Delta(G)$ and the graph onsite potential via $\Phi(G)$.

On the graph $G$, we introduce an operator-valued signal on each edge, analogous to the matrix-valued couplings used in non-Abelian lattices[27,28] and to vector diffusion maps for analysing high-dimensional datasets[29]. We define these edge operators with a set of $N_O \times N_O$ matrices $\{O_{p,q};\ 1 \leq p,q \leq N_G\}$ (Fig. 1b): elements of an $N_O^2$-dimensional Hilbert-Schmidt space $\mathcal{H}_O$. Collecting these operators, we construct a composite block operator $O = \sum_{p,q}(E_{p,q} \otimes O_{p,q}) \in \mathcal{H}_S \cong \mathcal{H}_G \otimes \mathcal{H}_O$, where $E_{p,q} = |p\rangle\langle q|$ is the standard basis operator of $\mathcal{H}_G$ associated with the $N_G$-dimensional state basis $\{|p\rangle\}$. Notably, $O$ fully encapsulates the edge operator information of $G$.

To capture the wave features of the operator signals, which are determined by the graph topology and vertex potential, we apply the graph Hamiltonian $H(G)$ to $O$, yielding $W(G) = (H(G) \otimes I_O)O \in \mathcal{H}_S$, where $I_O$ is the identity operator of $\mathcal{H}_O$. In mapping quantum dynamics onto graphs, we define the operator measure:

$$M(G) \triangleq \frac{1}{N_G}\mathrm{Tr}_G\left[W(G)\right] = \frac{1}{N_G}\mathrm{Tr}_G\left[\left(H(G)\otimes I_O\right)O\right], \tag{1}$$

which corresponds to an operator-valued graph energy expectation derived from the wave characteristics of $M$, obtained by averaging out the graph degrees of freedom through the partial trace over $\mathcal{H}_G$.

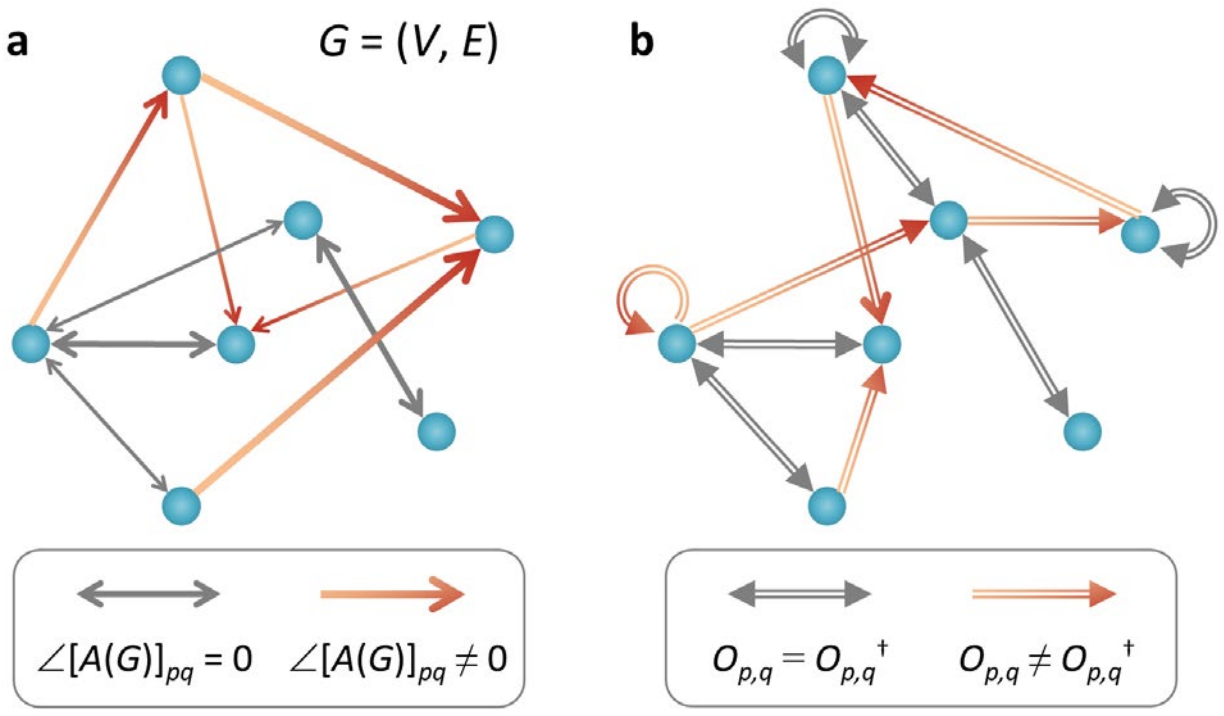

**Fig. 1. Schrödinger operator graphs. a,** Magnetic graph $G = (V, E)$ with $N_G$ vertices, where weighted edges depicted by their thicknesses represent $A(G)$. Grey and orange arrows denote vanishing and nonvanishing gauges, respectively. **b,** Operator-valued signals across the graph, where each edge $(p,q)$ carries an edge operator $O_{p,q}$. Grey and orange arrows indicate Hermitian and non-Hermitian signals, respectively. While $G$ is fully connected, only the dominant edges are shown for clarity.

## Lindbladian dynamics

We apply our framework to map $N$-dimensional Markovian quantum dynamics onto graph structures, examining the Lindblad QME[12,13] to cover both isolated and open quantum dynamics. Although our framework incorporates two degrees of freedom—graph architecture, including a vertex potential, and the distribution of an operator-valued signal—we pursue a universal description of quantum dynamics derived solely from the graph architecture. This requires a consistent assignment of operator-valued signals across the graph, which necessitates the use of the first standard form of the Lindblad QME rather than the reduced one[12]:

$$\frac{d\rho}{dt} = -\frac{i}{\hbar}[H,\rho] + \sum_{p,q} \gamma_{qp} \mathcal{D}_\rho(S_p, S_q), \tag{2}$$

with $\mathrm{D}_\rho(O_p,O_q) = O_p\rho O_q^\dagger - \{O_q^\dagger O_p, \rho\}/2$, where $\rho$ and $H$ denote the density operator and Hamiltonian, respectively, $\{S_p\}$ forms a basis for the $(N^2 - 1)$-dimensional subspace of traceless system-bath coupling operators acting on the system Hilbert space $\mathcal{H}$, and $\gamma_{pq}$ are the entries of the positive semidefinite (PSD) Kossakowski matrix $K$ with $[K]_{pq} = \gamma_{pq}$. Contrary to the reduced-form Lindblad QME, which requires altering both the decay rates and the jump operators for describing arbitrary systems, the first standard form allows for retaining an orthonormal basis $\{S_p\}$. This condition ensures that all degrees of freedom in nonunitary dynamics are fully encoded in $K$. In the analysis, we use the standard SU($N$) generator basis[30] for $\{S_p\}$, normalized as $\mathrm{Tr}[S_k^\dagger S_l] = \delta_{kl}$.

By adopting the vectorization for the Liouville space $\mathcal{L}_N$[31], Eq. (2) transforms into $d|\rho\rangle/dt = L|\rho\rangle$, where $|\rho\rangle \in \mathcal{L}_N$ denotes the superket of $\rho$, and $L = L_\mathrm{H} + L_\mathrm{L}$ is the Liouville superoperator, with

$$\begin{aligned} L_\mathrm{H} &= -\frac{i}{\hbar}\left\{\left[I_N \otimes H\right] - \left[H^T \otimes I_N\right]\right\}, \\ L_\mathrm{L} &= \sum_{p,q} \gamma_{qp}\left(S_q^* \otimes S_p - \frac{1}{2}\left(I_N \otimes S_q^\dagger S_p + S_p^T S_q^* \otimes I_N\right)\right), \end{aligned} \tag{3}$$

where $I_N$ is the $N \times N$ identity operator. While $L_\mathrm{H}$ describes the unitary evolution governed by the von Neumann equation, all nonunitary processes are governed by $L_\mathrm{L}$.

**Graph mapping**

We formulate Schrödinger operator graphs for unitary and nonunitary dynamics independently. First, for any Hermitian $H$ associated with $L_\mathrm{H}$ and a given state basis $\{|p\rangle\}$, we can always uniquely define an $N$-vertex magnetic graph $G_\mathrm{H}$ and its corresponding vertex potential $\Phi(G_\mathrm{H})$, yielding $H = \Delta(G_\mathrm{H}) + \Phi(G_\mathrm{H})$ (see Methods). By assigning an edge operator $O_{p,q}{}^\mathrm{H}$ from the $q$-th to the $p$-th vertex as

$$O_{p,q}{}^\mathrm{H} = -\frac{i}{\hbar}\left(I_N \otimes E_{p,q} - E_{q,p} \otimes I_N\right), \tag{4}$$

and constituting the block operator $O^{\mathrm{H}} = \sum_{p,q}(E_{p,q} \otimes O_{p,q}{}^{\mathrm{H}})$, the superoperator $L_{\mathrm{H}}$ can be expressed with our operator measure:

$$L_{\mathrm{H}} = \mathrm{Tr}_{\mathrm{G}}\left[\left(H(G_{\mathrm{H}}) \otimes I_N\right)O^{\mathrm{H}}\right] \triangleq NM^{\mathrm{H}}(G_{\mathrm{H}}). \tag{5}$$

Similarly, we introduce the $(N^2 - 1)$-vertex magnetic graph $G_{\mathrm{L}}$ that derives the Kossakowski matrix $K = \Delta(G_{\mathrm{L}}) + \Phi(G_{\mathrm{L}})$ for a given operator basis $\{S_p\}$ (see Methods). By defining the $(p,q)$ edge operator as

$$O_{p,q}{}^{\mathrm{L}} = S_q{}^{*} \otimes S_p - \frac{1}{2}\left(I_N \otimes S_q{}^{\dagger}S_p + S_p{}^{T}S_q{}^{*} \otimes I_N\right), \tag{6}$$

and its composite form $O^{\mathrm{L}} = \sum_{p,q}(E_{p,q} \otimes O_{p,q}{}^{\mathrm{L}})$, the nonunitary evolution is governed by the following operator measure:

$$L_{\mathrm{L}} = \mathrm{Tr}_{\mathrm{G}}\left[\left(K(G_{\mathrm{L}}) \otimes I_{N^2}\right)O^{\mathrm{L}}\right] = \mathrm{Tr}_{\mathrm{G}}\left[W\right] \triangleq \left(N^2 - 1\right)M^{\mathrm{L}}(G_{\mathrm{L}}). \tag{7}$$

Consequently, our framework provides a mathematically rigorous mapping of Markovian quantum dynamics onto the graphs $G_{\mathrm{H}}$ and $G_{\mathrm{L}}$, as $d|\rho\rangle/dt = [NM^{\mathrm{H}}(G_{\mathrm{H}}) + (N^2 - 1)M^{\mathrm{L}}(G_{\mathrm{L}})]|\rho\rangle$. We note that the signals at the $p$-th vertex of $G_{\mathrm{H}}$ and $G_{\mathrm{L}}$ can be interpreted as $|p\rangle$ and $S_p$, respectively, which dictate the edge signals via Eqs. (4) and (6). Once the bases are fixed, the entire open-system dynamics are completely characterized solely by the topologies of the two graphs, $G_{\mathrm{H}}$ and $G_{\mathrm{L}}$, and their vertex potentials, $\Phi(G_{\mathrm{H}})$ and $\Phi(G_{\mathrm{L}})$.

**Open quantum Rabi graphs**

As an illustrative application of our framework, we consider the open QRM (Fig. 2a). The dipole-gauge QRM Hamiltonian for a single photonic mode of frequency $\omega_0$, with vector potential $\mathbf{A} = \mathbf{x}A_0(a + a^{\dagger})$, is[32-35]

$$H = \frac{E_{10}}{2}\sigma_z + \hbar\omega_0 a^{\dagger}a + \hbar\omega_0\eta^2 - i\hbar\omega_0\eta\sigma_x(a - a^{\dagger}), \tag{8}$$

where $a$ and $a^\dagger$ are the annihilation and creation operators, respectively, $\sigma_{x,y,z}$ denote the Pauli matrices, $E_{10}$ is the qubit level splitting, and $\eta = A_0 d/\hbar$ denotes the light-qubit coupling strength, with $d$ being the differential dipole moment. To describe dissipation in a form valid beyond the strong coupling regime[33,34,36], we treat the system-bath coupling in the dressed-state picture[37]. Including cavity relaxation at rate $\kappa$, qubit relaxation at rate $\gamma$, and qubit dephasing at rate $\gamma_\varphi$, the Lindblad QME becomes

$$\begin{aligned}\frac{d\rho}{dt} &= -\frac{i}{\hbar}[H,\rho] \\ &+ \sum_{p,q>p}\left(\kappa\left|\langle p|\left(a+a^\dagger\right)|q\rangle\right|^2 + \gamma\left|\langle p|\sigma_x|q\rangle\right|^2\right)\mathcal{D}_\rho(\sigma_{pq}) \\ &+ \sum_{p\neq q}\frac{\gamma_\varphi}{2}\left|\langle p|\sigma_z|q\rangle\right|^2\mathcal{D}_\rho(\sigma_{pq}) + \mathcal{D}_\rho\Big(\sum_p\sqrt{\frac{\gamma_\varphi}{2}}\langle p|\sigma_z|p\rangle\sigma_{pp}\Big),\end{aligned} \tag{9}$$

where $\{|p\rangle\}$ is the eigenbasis of $H$ sorted in nondecreasing order of eigenenergy, $\sigma_{pq} = |p\rangle\langle q|$, and $\mathrm{D}_\rho(O) \triangleq \mathrm{D}_\rho(O,O)$.

Figures 2b and 2c show $G_\mathrm{H}$ and $G_\mathrm{L}$, respectively, which describe the open QRM in the weak coupling regime and experimentally feasible dissipative parameters $(\kappa, \gamma, \gamma_\varphi)$[36,37]. The corresponding potentials $\Phi(G_\mathrm{H})$ and $\Phi(G_\mathrm{L})$ are encoded in the vertex colours. Notably, $G_\mathrm{H}$ coincides with the Fock-state lattice[7]—a bipartite graph characterized by a vertex potential that increases linearly with the photon number. By contrast, the nonunitary graph $G_\mathrm{L}$ generalizes this lattice-based description for isolated systems to open quantum dynamics by encoding the contributions of coupling operators to magnetic graph edges. Remarkably, the Liouville spectrum $\lambda$ in Fig. 2d exhibits a nearly symmetric, regular, and rhombus-shaped distribution. Although the spectral boundary is reminiscent of the lemon-shaped spectra of fully connected random Lindbladian operators governed by Gaussian statistics[15], the discretized nature of the spectrum reflects the underlying quasi-regular graph structures of $G_\mathrm{H}$ and $G_\mathrm{L}$.

To gain insight into the relationship between the QRM graphs and Liouville spectra, we examine each dissipative process separately: cavity relaxation (Fig. 2e,f), qubit relaxation (Fig. 2g,h), and qubit dephasing (Fig. 2i,j). First, the relaxation and dephasing processes are distinguished by their edge phases: the relaxation graphs with $\pm\pi/2$ phases and the dephasing graph with 0 or $\pi$ phases, which reflect the dominance and suppression of level transitions in each process, respectively. The cavity relaxation graph is quasi-identical to the full dissipation case with the densest connectivity among the individual processes, identifying cavity loss as the dominant contributor to the overall spectrum. By contrast, the qubit relaxation graph is considerably sparser, and its spectrum collapses into discrete values of Re[$\lambda$] with pronounced degeneracies, reflecting the limited number of dressed-state transitions. The dephasing graph is the sparsest, because dephasing affects mainly coherences with negligible contributions to population transitions in the weak coupling regime. Notably, its topology features a pronounced hub structure—a small number of vertices concentrate the majority of edge connections—leading to fewer distinct decay rates than in the qubit relaxation and the consequent concentration around a dominant value of Re[$\lambda$]. These results show that distinct dissipation processes map onto distinct graph motifs, exhibiting diverse spectral signatures.

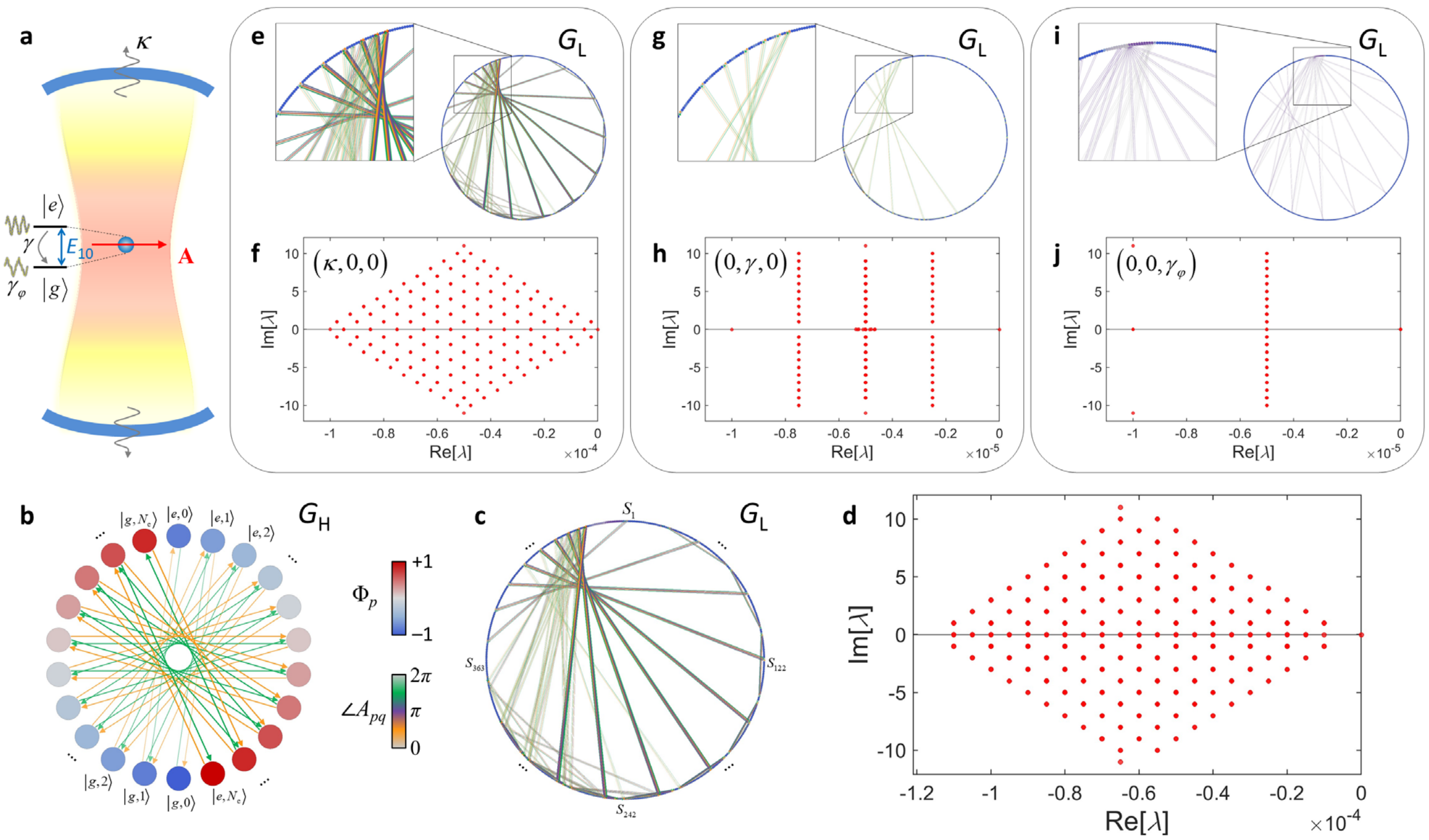

**Fig. 2. Open quantum Rabi graphs. a,** Schematic of the open QRM. **b,c,** $G_H$ (**b**) and $G_L$ (**c**) for $\eta$ = $10^{-3}$ and ($\kappa$, $\gamma$, $\gamma_\varphi$) = ($10^{-5}$, $10^{-5}$, $10^{-5}$). **d,** Corresponding Liouville spectrum $\lambda$. **e-j,** $G_L$ and spectra for ($\kappa$, $\gamma$, $\gamma_\varphi$) = ($10^{-5}$, 0, 0) (**e,f**), ($\kappa$, $\gamma$, $\gamma_\varphi$) = (0, $10^{-5}$, 0) (**g,h**), and ($\kappa$, $\gamma$, $\gamma_\varphi$) = (0, 0, $10^{-5}$) (**i,j**). For clarity, only edges with $|A_{pq}|$ > max($\{|A_{pq}|\}$)/1000 are shown. Edge thickness and opacity encode $|A_{pq}|$, edge colour encodes $\angle A_{pq}$, and vertex colour encodes normalized $\Phi_p$. The Fock basis is truncated at $N_{max}$ = 10. $\hbar$ = 1 and $E_{10} = \hbar\omega_0$ = 1.

## Ultrastrong coupling

We investigate the graph-theoretic interpretation of the transition from the weak-coupling to USC regimes of the QRM. Figures 3a-c show $G_L$ in the USC regime at $\eta$ = 0.1 for the total dissipation (Fig. 3a), cavity relaxation only (Fig. 3b), and qubit dephasing only (Fig. 3c) (see Supplementary Note S1 for qubit relaxation). All cases exhibit a markedly denser architecture than those in Fig. 2, simultaneously exhibiting an enhanced contribution of dephasing to graph topology, in accordance with the analysis in Methods. To quantify this increasing connectivity, we evaluate the Fiedler value $\lambda_1$ (see Methods), which determines a lower bound of the cost in separating a large phase-coherent subgraph according to the generalized Cheeger inequality[24]. As shown in Fig. 3d, $\lambda_1$ increases linearly with $\eta$, establishing that the USC regime is accompanied by a pronounced increase in the connectivity between system-bath coupling operators. Another notable feature is revealed by the vertex degree, $d_p(G_{\mathrm{L}}) = \sum_q |[A(G_{\mathrm{L}})]_{pq}|$. Figure 3e shows that, as $\eta$ increases, vertices with large degree progressively emerge, leading to the formation of hub-like vertices. The emergence of these hubs is in line with the graph architecture described in Figs. 2i and 3c, confirming the enhanced influence of qubit dephasing in the USC regime.

Along with these graph-theoretic signatures, the spectrum becomes increasingly continuous as $\eta$ increases (Figs. 3f,3g; Supplementary Note S1 for individual processes). This trend is consistent with the enhanced graph connectivity—analogous to the statistically filled universal spectrum of fully connected graphs[15]. Consequently, the results demonstrate that our framework successfully captures the evolution of the QRM spectrum from the weak-coupling to USC regime.

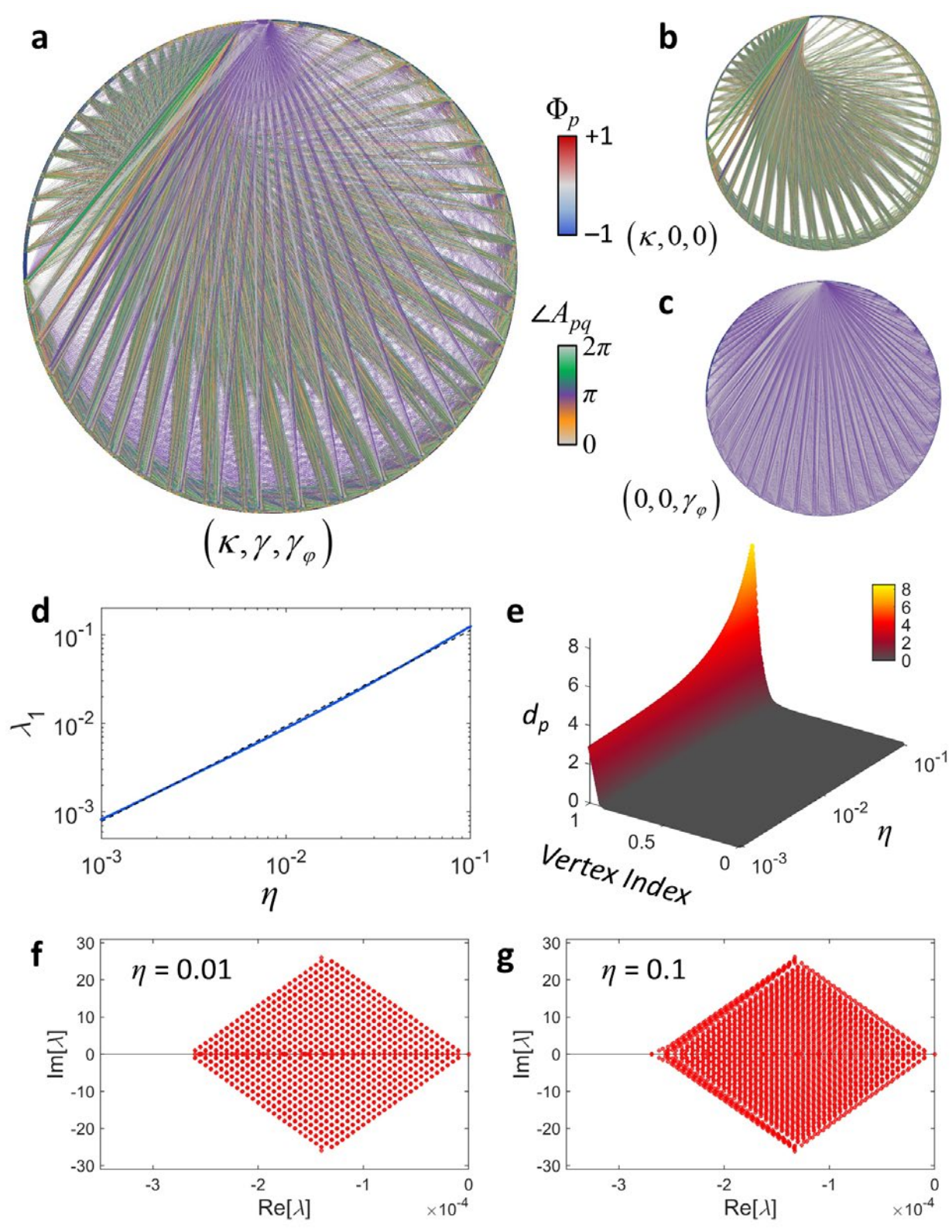

**Fig. 3. USC open quantum Rabi graphs. a-c,** $G_L$ for $(\kappa, \gamma, \gamma_\varphi) = (10^{-5}, 10^{-5}, 10^{-5})$ (**a**), $(10^{-5}, 0, 0)$ (**b**), and $(0, 0, 10^{-5})$ (**c**). **d,** $\eta$-dependent $\lambda_1$ (blue solid line) and its fitting $\lambda_1 = 1.35 \times \eta^{1.08}$. **e,** Sorted degree distributions versus $\eta$. **f,g,** Liouville spectra for $\eta = 0.01$ (**f**) and 0.1 (**g**). $N_{max} = 25$ to cover the USC regime.

## Graph sparsification

To validate our graph-theoretic framework as a cost-efficient abstraction of quantum dynamics, we devise edge pruning of our Schrödinger operator graphs, focusing on the sparsification of $G_L$ for $K = \Delta(G_L) + \Phi(G_L)$. We prune edges of small magnitude by setting $[A(G_L)]_{pq} = 0$ when $|[A(G_L)]_{pq}|/\max[\{|[A(G_L)]_{pq}|\}] \leq \chi$ for the pruning threshold $\chi$. For the pruned graph $G_L^P$, we compensate the degree deficit $\delta_p = d_p(G_L) - d_p(G_L^P) \geq 0$ through the vertex potential, as $[\Phi(G_L^P)]_{pp} = [\Phi(G_L)]_{pp} + \delta_p$ (Fig. 4a), thereby stabilizing the PSD structure of the pruned Kossakowski matrices, $K^P = \Delta(G_L^P) + \Phi(G_L^P)$. Figures 4b-4d show the resulting spectra of the USC QRM.

Remarkably, the overall spectral distribution remains robust even at $\chi = 0.6$: the characteristic rhombus-shaped boundary and the interior spectral density are largely preserved. Deviations become apparent mainly at the spectral periphery, while the bulk of the spectrum retains its structure.

To quantify the pruning performance, we compare the result with the direct sparsification of the Kossakowski matrix $K$: setting $[K]_{pq} = 0$ when $|[K]_{pq}|/\max[\{|[K]_{pq}|\}] \leq \chi$, analogous to the graph pruning. For a fair comparison, the diagonal entries of the pruned matrix $K'$ are corrected by compensating for the degree deficit $\delta_p = d_p^K - d_p^{K'} \geq 0$ as $[K'']_{pp} = [K']_{pp} + \delta_p$, where $d_p^K = \sum_q |[K]_{pq}|$. To ensure the PSD condition of the Kossakowski matrix, we utilize the scalar-shifted one, $K^{\mathrm{F}} = K'' + \alpha I$, for calculating the eigenspectrum of the sparsified system, where $\alpha = \max(0, -\lambda_{\min}(K''))$ for the minimum eigenvalue $\lambda_{\min}(K'')$ of $K''$. For $K^{\mathrm{P}}$ and $K^{\mathrm{F}}$, we obtain the corresponding Liouville spectra $\Lambda^{\mathrm{P}} = \{\lambda_n^{\mathrm{P}}\}$ and $\Lambda^{\mathrm{F}} = \{\lambda_n^{\mathrm{F}}\}$ for varying $\chi$.

Figure 4e shows the distances of $\Lambda^{\mathrm{P}}$ and $\Lambda^{\mathrm{F}}$ from the original spectrum $\Lambda$, measured by the Chamfer distance $D(\Lambda^X, \Lambda)$ with $X \in \{\mathrm{P}, \mathrm{F}\}$ (see Methods). Across the entire range of $\chi$, graph pruning achieves a substantially smaller Chamfer distance than that of the matrix sparsification, demonstrating that graph pruning enables superior abstraction of open quantum dynamics (see Supplementary Note S2 for extended results). These results also indicate that the dominant spectral features are governed by a relatively small subset of strong graph edges and the vertex potential, which together form the backbone of our framework.

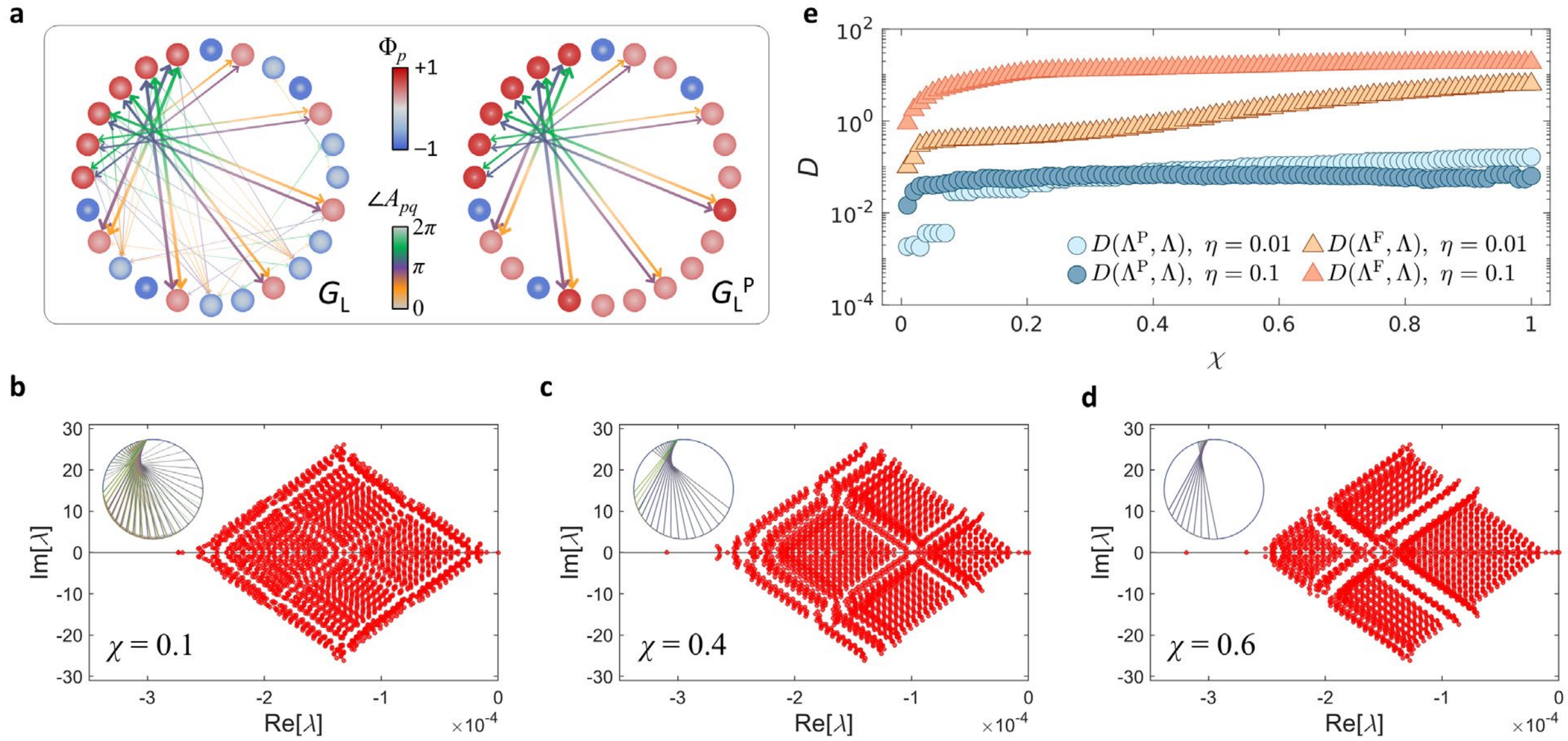

**Fig. 4. Graph pruning. a,** Schematic of graph pruning. **b-d,** Liouville spectra and the corresponding $G_L{}^P$ for $\chi = 0.1$ (**b**), 0.4 (**c**), and 0.6 (**d**) at $\eta = 0.1$. **e,** Chamfer distance versus $\chi$ for graph pruning, $D(\Lambda^P, \Lambda)$, and matrix sparsification, $D(\Lambda^F, \Lambda)$, at $\eta = 0.01$ and 0.1. All other parameters are the same as those in Fig. 3 in the main text.

## Graph neural networks

The observed robustness of our graph-theoretic framework in Fig. 4 inspires the data-driven inference of open quantum dynamics based on the efficient operator-graph structures. As a representative example, we develop a GCN[38] for the regression of the Liouville spectra of random open QRMs (Fig. 5a), using $G_H$ and $G_L$ as inputs, where the graphs have the features of $\Phi(G_H)$ and $\Phi(G_L)$ at their vertices, respectively. Considering the dimensionalities of $G_H$ and $G_L$, two and four GCN layers are implemented in the subnetworks of $G_H$ and $G_L$, respectively. After the GCN layers, the extracted features are mixed via fully connected layers. The overall network is trained on the eigenspectrum distributions of random open QRMs (Supplementary Note S3 for details).

Figures 5b and 5c show the GCN performances: the Chamfer distances of 1000 random open QRMs for varying $\chi$ and the optimal example, respectively (see Supplementary Note S4 for extended data). Remarkably, although the pristine graphs with $\chi = 0$ include the complete information of open QRMs, superior performance is obtained with enhanced pruning up to $\chi = 0.75$ (Fig. 5b). However, further pruning beyond $\chi = 0.75$ results in the degradation of the regression performance.

This observation can be elucidated by the competition between the improved learning of the GCN and the information loss of open quantum dynamics. Owing to the fully connected, weighted configuration of the pristine $G_L$, negligible edges in $G_L$ can act as noise that hinders the training of the GCN[39]. Therefore, pruning these weakly weighted edges by increasing $\chi$ substantially improves the training efficiency of the GCN, as demonstrated by the dramatically reduced training time required to reach the performance level of the pristine case, $\chi = 0$ (Fig. 5d; Supplementary Note S5). Although the loss of information from pruning hub edges at $\chi > 0.75$ outweighs this noise-reduction benefit, the better performance even at $\chi = 1$ than in the pristine case demonstrates the substantial role of another graph component: the vertex potential.

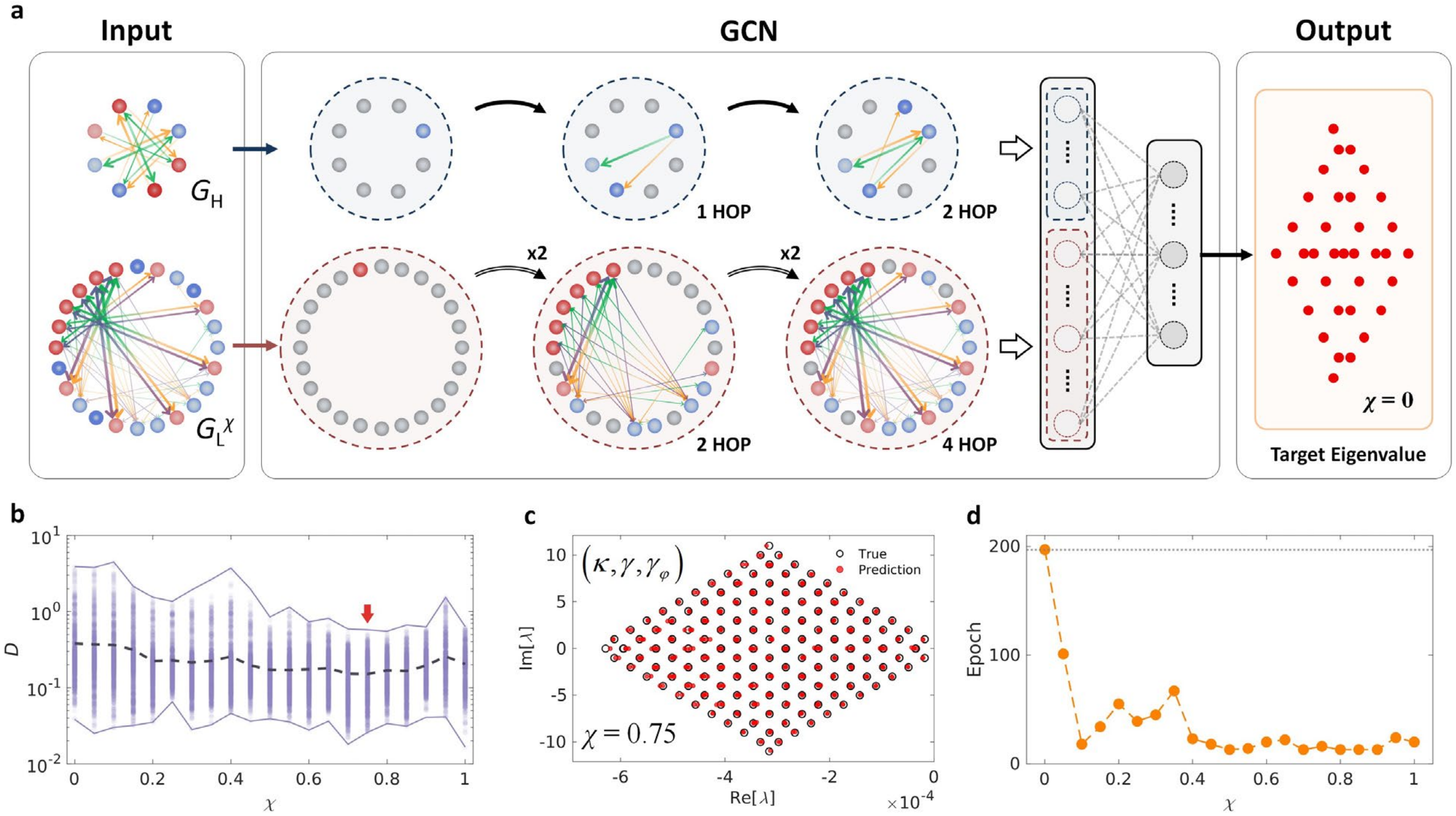

**Fig. 5. Schrödinger-operator graph neural networks. a,** Schematic of the GCN, where 'HOP' denotes the message-passing range. **b,** Chamfer distance between the true and predicted eigenspectra for varying $\chi$. Dots and the dashed line denote the results of random realizations and their averages, respectively. The solid lines represent the minimum and maximum values of $D$ for 1000 realizations at each $\chi$. The red arrow depicts $\chi = 0.75$ yielding the lowest average. **c,** The optimal case at $\chi = 0.75$ for the realization $(\kappa, \gamma, \gamma_\varphi) = (6.21\times10^{-5}, 7.43\times10^{-6}, 1.27\times10^{-6})$. **d,** The training epoch at which the pristine-graph ($\chi = 0$) performance level is reached. $\eta = 0.001$ in all cases.

## Discussion

In our example study, we establish a forward mapping from a given open quantum system—the open QRM—onto its Schrödinger operator graphs. A largely unexplored complementary direction is the inverse question: which classes of quantum dynamics are supported by prescribed graph topologies. To explore the universal spectral features of Markovian quantum systems[15,23], systematically surveying graph families[40]—the Erdős–Rényi model, small-world, scale-free, and regular graphs—could reveal how distinct topological motifs yield characteristic Liouville spectra. Such a topology-to-dynamics correspondence would link graph-theoretic metrics, such as degree distributions, centrality, path lengths, and clustering, to universal features of dissipative spectra.

Our GCN also currently performs a forward inference, predicting Liouville spectra from the graph representation of a given quantum system. Given the growing success of neural networks in inverse design[41], a natural extension is to reverse this flow: to infer the underlying quantum system from experimental measurements, and ultimately to design systems that realize targeted quantum observables. Whereas open quantum dynamics has conventionally been addressed through forward, theory-driven modelling[37], a learning-based inverse framework built on graph abstraction would offer a novel route to handling computationally expensive, high-dimensional quantum systems. The recently proposed gradient-based optimization of quantum systems[42,43] and graph architectures[44] also suggests alternative route to such inverse-design tasks.

In conclusion, we have proposed a graph-theoretic framework and demonstrated its application to Markovian quantum dynamics. This formulation describes open quantum dynamics in terms of two magnetic graphs with vertex potentials, under a mathematically rigorous correspondence with the Lindblad QME. Within this framework, distinct quantum phenomena can be classified by graph topology, as illustrated by the open QRM across the weak-coupling and USC regimes. Moreover, the physically allowed graph sparsification via edge pruning preserves the critical features of the pristine dynamics while improving both the accuracy and the training efficiency of the GCN. We expect that our framework, by revealing how the connectivity of system-bath coupling governs open quantum dynamics, will stimulate a graph-theoretic understanding of quantum phenomena.

## Methods

**Magnetic graph Laplacian.** For a magnetic graph $G$, its complex-valued adjacency matrix $A(G)$ is Hermitian with a vanishing diagonal, while each off-diagonal entry encodes the weighted coupling between two vertices. The corresponding degree matrix is defined by $[D(G)]_{pq} = d_p\delta_{pq} = \delta_{pq}\sum_q|[A(G)]_{pq}|$, so that each diagonal entry collects the total incident edge weight at the corresponding vertex. The magnetic Laplacian for the Schrödinger operator graph is the unnormalized matrix $\Delta(G) = D(G) - A(G)$, which is also Hermitian.

When characterizing the graph connectivity using the Fiedler value as shown in Fig. 3, we employ the normalized magnetic Laplacian, $\Delta_{\mathrm{norm}}(G) = D(G)^{-1/2}\Delta(G)D(G)^{-1/2}$, to remove the trivial dependence on local degree scale and thereby enable a fair comparison of connectivity across graphs with different degree distributions. Among the eigenvalues of $\Delta(G)_{\mathrm{norm}}$, denoted by $\{\lambda_j\}$ and sorted in nondecreasing order, we focus on the Fiedler value—the smallest positive eigenvalue of $\Delta_{\mathrm{norm}}(G_{\mathrm{L}})$: $\lambda_1 = \min\{\lambda_j > 0\}$. According to the generalized Cheeger inequality[24], this quantity provides a spectral measure of global connectivity: a larger $\lambda_1$ indicates stronger coherent interconnection, or equivalently, a higher cost of separating a large phase-coherent subgraph.

**Graph uniqueness.** Consider a Hermitian matrix $Q$. We assume that $Q$ can be decomposed as $Q = \Delta(G) + \Phi(G)$ for a magnetic graph $G$. Because $\Phi(G)$ is diagonal, the off-diagonal entries of $Q$ uniquely determine the adjacency matrix of $G$ through $[A(G)]_{pq} = (\delta_{pq} - 1)[Q]_{pq}$. The degree matrix and the vertex potential matrix are then given by $[D(G)]_{pq} = \delta_{pq}\sum_{r\neq p}|[Q]_{pr}|$ and $[\Phi(G)]_{pq} = \delta_{pq}(Q_{pp} - \sum_{r\neq p}|[Q]_{pr}|)$, respectively. Therefore, both the graph $G$ and the vertex potential $\Phi(G)$ are uniquely determined once $Q$ is given. Because the Hamiltonian matrix $H$ and the Kossakowski matrix $K$ are both Hermitian, their associated graphs are determined uniquely as well.

**Enhanced dephasing with increasing coupling.** In the weak coupling regime ($\eta \ll 1$), the dressed eigenstates of Eq. (8) remain perturbatively close to the eigenstates $\{|g, n\rangle \triangleq |n_0^-\rangle, |e, n\rangle \triangleq |n_0^+\rangle | n \in \mathbb{N}_0\}$ of the uncoupled Hamiltonian

$$H_0 = \frac{E_{10}}{2}\sigma_z + \hbar\omega_0 a^\dagger a. \tag{10}$$

For the eigenvalues $E_{n0}^{\pm} = n\hbar\omega_0 \pm E_{10}/2$ from $H_0|n_0^{\pm}\rangle = E_{n0}^{\pm}|n_0^{\pm}\rangle$, we assume nondegeneracy by setting $E_{10} < \hbar\omega_0$. When we set $V = -i\hbar\omega_0\sigma_x(a - a^\dagger)$, the perturbed states become

$$\left|n^{\pm}\right\rangle = \left|n_0^{\pm}\right\rangle + \eta\left(\sum_{r\neq n}\frac{\left\langle r_0^{\pm}\middle|V\middle|n_0^{\pm}\right\rangle}{E_{n0}^{\pm} - E_{r0}^{\pm}}\left|r_0^{\pm}\right\rangle + \sum_{s}\frac{\left\langle s_0^{\mp}\middle|V\middle|n_0^{\pm}\right\rangle}{E_{n0}^{\pm} - E_{s0}^{\mp}}\left|s_0^{\mp}\right\rangle\right) + O(\eta^2). \tag{11}$$

Using these perturbed states, the coefficient of the dephasing process becomes

$$\left\langle p^{\pm}\middle|\sigma_z\middle|q^{\pm}\right\rangle = \pm\delta_{pq} + O(\eta^2), \tag{12}$$

$$\left\langle p^{\mp}\middle|\sigma_z\middle|q^{\pm}\right\rangle = \pm 2\eta\frac{\left\langle p_0^{\mp}\middle|V\middle|q_0^{\pm}\right\rangle}{E_{p0}^{\mp} - E_{q0}^{\pm}} + O(\eta^2), \tag{13}$$

In contrast to other relaxation processes, the dephasing in the weak coupling regime is dominated by the diagonal components in Eq. (12), whose off-diagonal contributions scale as $O(\eta^2)$ according to Eq. (13). Because the diagonal terms affect only coherences without inducing population transitions, their contribution is confined to Im[$\lambda$], rendering the dephasing contribution to Re[$\lambda$] negligible at small $\eta$. However, as $\eta$ increases, the off-diagonal matrix elements in Eq. (13) increase linearly, introducing dephasing-induced interlevel transitions and a consequent contribution to Re[$\lambda$].

**Chamfer distance.** For each point in one set, the Chamfer distance finds the Euclidean distance to its nearest point in the other set. Averaging these nearest-neighbour distances over each set and summing the two directional terms yields measure of the difference between the two point sets.

We represent each eigenvalue $\lambda$ as a normalized point $\lambda_{norm}$ = (Re[$\lambda$]/$\sigma_{Re}$, Im[$\lambda$]/$\sigma_{Im}$), where $\sigma_{Re}$ and $\sigma_{Im}$ are the standard deviations of the real and imaginary parts of the original spectrum $\Lambda = \{\lambda_n\}$, placing the real and imaginary axes—which differ by several orders of magnitude—on a common scale. For a pruned spectrum $\Lambda^X = \{\lambda_n^X\}$ and the original spectrum $\Lambda$, the Chamfer distance is

$$D(\Lambda^X, \Lambda) = \frac{1}{|\Lambda^X|} \sum_{\lambda' \in \Lambda^X} \min_{\lambda \in \Lambda} \left\| \lambda_{norm}' - \lambda_{norm} \right\| + \frac{1}{|\Lambda|} \sum_{\lambda \in \Lambda} \min_{\lambda' \in \Lambda^X} \left\| \lambda_{norm}' - \lambda_{norm} \right\|, \tag{14}$$

where ||·|| is the Euclidean norm; a smaller $D$ indicates better preservation of the original spectrum.

## Data availability

Data used in the current study are available from the corresponding authors upon request and can also be obtained by running the shared codes at 10.5281/zenodo.19777226 in the Zenodo.

## Code availability

All original code has been deposited at Zenodo and is publicly available at 10.5281/zenodo.19777226.

## Acknowledgements

We acknowledge financial support from the National Research Foundation of Korea (NRF) through the Innovation Research Center (No. RS-2024-00413957), Core Research Grants (No. RS-2026-25469085), Pilot and Feasibility Grants (No. RS-2025-19912971), and Young Researcher Program (No. RS-2025-00552989), all funded by the Korean government. This work was supported by the BK21 FOUR program of the Education and Research Program for Future

ICT Pioneers in 2026, through Seoul National University. We also acknowledge an administrative support from SOFT foundry institute.

## Author Contributions

S.Y., X.P., and N.P. conceived the idea. K.K., S.Y., X.P., and N.P. developed the theoretical tool and performed the numerical analysis. K.K., D.L., and S.P. examined the theoretical and numerical analysis. X.P., N.P., and S.Y. supervised the findings of this work. All authors discussed the results and wrote the final manuscript.

## Competing Interests

The authors have no conflicts of interest to declare.

## Additional information

**Correspondence and requests for materials** should be addressed to X.P., N.P., or S.Y.

## Figure Legends

**Fig. 1. Schrödinger operator graphs. a,** Magnetic graph $G = (V, E)$ with $N_G$ vertices, where weighted edges depicted by their thicknesses represent $A(G)$. Grey and orange arrows denote vanishing and nonvanishing gauges, respectively. **b,** Operator-valued signals across the graph, where each edge $(p,q)$ carries an edge operator $O_{p,q}$. Grey and orange arrows indicate Hermitian and non-Hermitian signals, respectively. While $G$ is fully connected, only the dominant edges are shown for clarity.

**Fig. 2. Open quantum Rabi graphs. a,** Schematic of the open QRM. **b,c,** $G_H$ (**b**) and $G_L$ (**c**) for $\eta = 10^{-3}$ and $(\kappa, \gamma, \gamma_\varphi) = (10^{-5}, 10^{-5}, 10^{-5})$. **d,** Corresponding Liouville spectrum $\lambda$. **e-j,** $G_L$ and spectra for $(\kappa, \gamma, \gamma_\varphi) = (10^{-5}, 0, 0)$ (**e,f**), $(\kappa, \gamma, \gamma_\varphi) = (0, 10^{-5}, 0)$ (**g,h**), and $(\kappa, \gamma, \gamma_\varphi) = (0, 0, 10^{-5})$ (**i,j**). For clarity, only edges with $|A_{pq}| > \max(\{|A_{pq}|\})/1000$ are shown. Edge thickness and opacity encode $|A_{pq}|$, edge colour encodes $\angle A_{pq}$, and vertex colour encodes normalized $\Phi_p$. The Fock basis is truncated at $N_{\max} = 10$. $\hbar = 1$ and $E_{10} = \hbar\omega_0 = 1$.

**Fig. 3. USC open quantum Rabi graphs. a-c,** $G_L$ for $(\kappa, \gamma, \gamma_\varphi) = (10^{-5}, 10^{-5}, 10^{-5})$ (**a**), $(10^{-5}, 0, 0)$ (**b**), and $(0, 0, 10^{-5})$ (**c**). **d,** $\eta$-dependent $\lambda_1$ (blue solid line) and its fitting $\lambda_1 = 1.35 \times \eta^{1.08}$. **e,** Sorted degree distributions versus $\eta$. **f,g,** Liouville spectra for $\eta = 0.01$ (**f**) and 0.1 (**g**). $N_{\max} = 25$ to cover the USC regime.

**Fig. 4. Graph pruning. a,** Schematic of graph pruning. **b-d,** Liouville spectra and the corresponding $G_L^P$ for $\chi = 0.1$ (**b**), 0.4 (**c**), and 0.6 (**d**) at $\eta = 0.1$. **e,** Chamfer distance versus $\chi$ for graph pruning, $D(\Lambda^P, \Lambda)$, and matrix sparsification, $D(\Lambda^F, \Lambda)$, at $\eta = 0.01$ and 0.1. All other parameters are the same as those in Fig. 3 in the main text.

**Fig. 5. Schrödinger-operator graph neural networks. a,** Schematic of the GCN, where 'HOP' denotes the message-passing range. **b,** Chamfer distance between the true and predicted eigenspectra for varying $\chi$. Dots and the dashed line denote the results of random realizations and their averages, respectively. The solid lines represent the minimum and maximum values of $D$ for 1000 realizations at each $\chi$. The red arrow depicts $\chi = 0.75$ yielding the lowest average. **c,** The optimal case at $\chi = 0.75$ for the realization $(\kappa, \gamma, \gamma_\varphi) = (6.21\times10^{-5}, 7.43\times10^{-6}, 1.27\times10^{-6})$. **d,** The training epoch at which the pristine-graph ($\chi = 0$) performance level is reached. $\eta = 0.001$ in all cases.

# Supplementary Information for "Mapping open quantum dynamics onto graphs"

Kyuho Kim[1], Dayeong Lee[1], Seungkyun Park[1,2,4], Xianji Piao[3§], Namkyoo Park[2†], and Sunkyu Yu[1*]

[1]Intelligent Wave Systems Laboratory, Department of Electrical and Computer Engineering, Seoul National University, Seoul 08826, Korea

[2]Photonic Systems Laboratory, Department of Electrical and Computer Engineering, Seoul National University, Seoul 08826, Korea

[3]Wave Engineering Laboratory, School of Electrical and Computer Engineering, University of Seoul, Seoul 02504, Korea

[4]KAIST InnoCORE PICORE Center, Korea Advanced Institute of Science and Technology, Daejeon, 34141, Korea

E-mail address for correspondence: [§]piao@uos.ac.kr, [†]nkpark@snu.ac.kr, [*]sunkyu.yu@snu.ac.kr

## Supplementary Note S1. Extended analysis of the USC regime

Figure S1 shows $G_L$ in the USC regime at $\eta = 0.1$ for qubit relaxation only. Notably, $G_L$ of the qubit relaxation also shows a markedly denser architecture than that in Fig. 2g similarly to the other dissipation processes. Figures S1b-d show the resulting spectra from individual dissipation processes, which correspond to the USC counterparts of Figs. 2f,h,j in the main text. The results show that while cavity relaxation still dominates the spectral features in this regime, the discreteness of both the qubit relaxation and qubit dephasing spectra becomes mitigated as interlevel transitions increase.

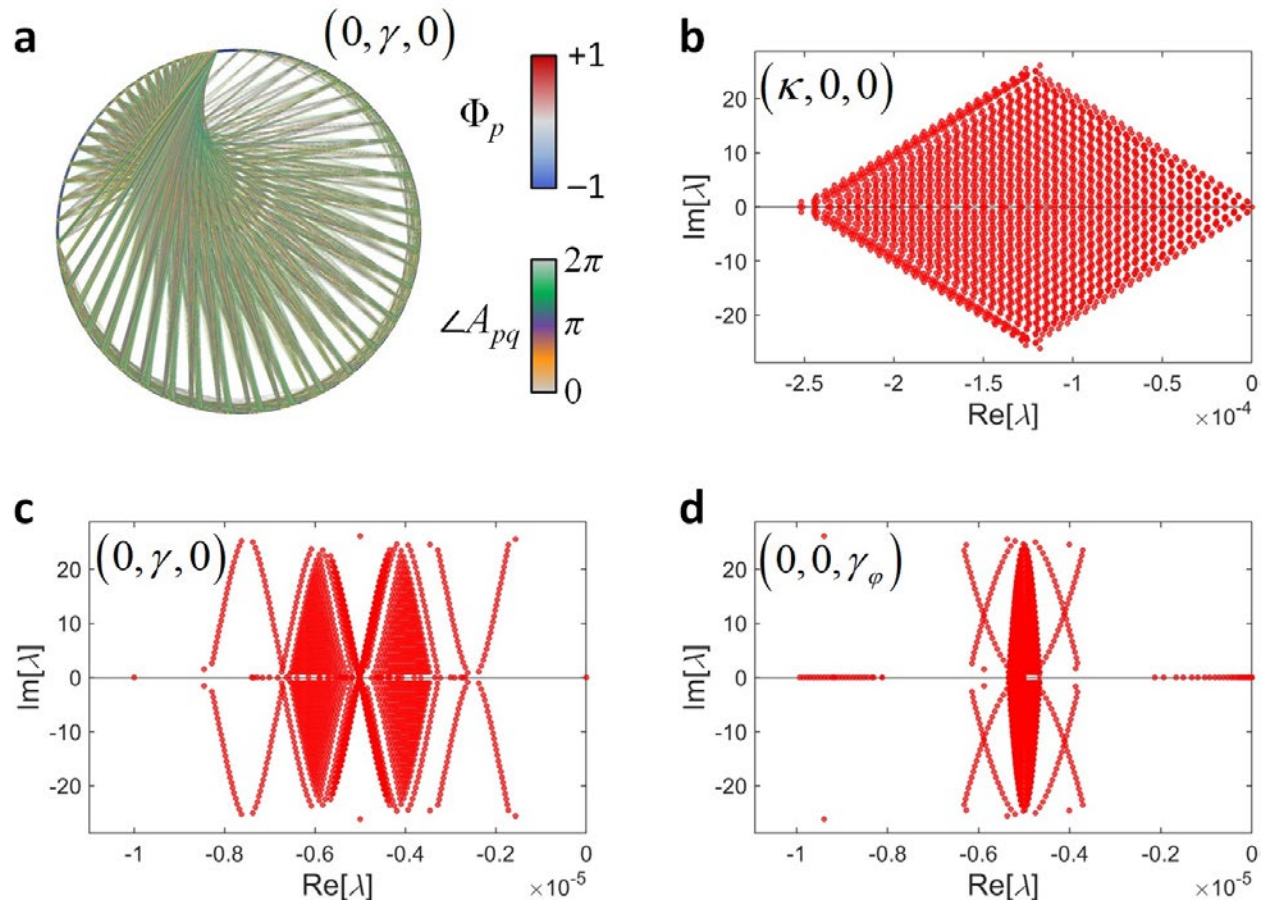

**Supplementary Figure S1. Extended analysis of the operator graph and spectra in the USC regime.** **a**, $G_L$ for $(0, \gamma, 0)$. **b-d**, Liouville spectra for $(\kappa, 0, 0)$ (**b**), $(0, \gamma, 0)$ (**c**), and $(0, 0, \gamma_\varphi)$ (**d**). All the other parameters are the same as those in Fig. 3.

**Supplementary Note S2. Extended analysis of graph pruning**

Figure S2 demonstrates the superior performance of graph pruning over matrix pruning across different coupling strengths. Reflecting the highly deformed Liouville spectra in the matrix pruning case even with relatively small $\chi$ (Figs. S2a–d), graph pruning exhibits a Chamfer distance nearly two orders of magnitude smaller than that of matrix pruning (Fig. S2e), demonstrating that graph pruning enables significantly better preservation of the spectral features of the QRM regardless of the coupling strength.

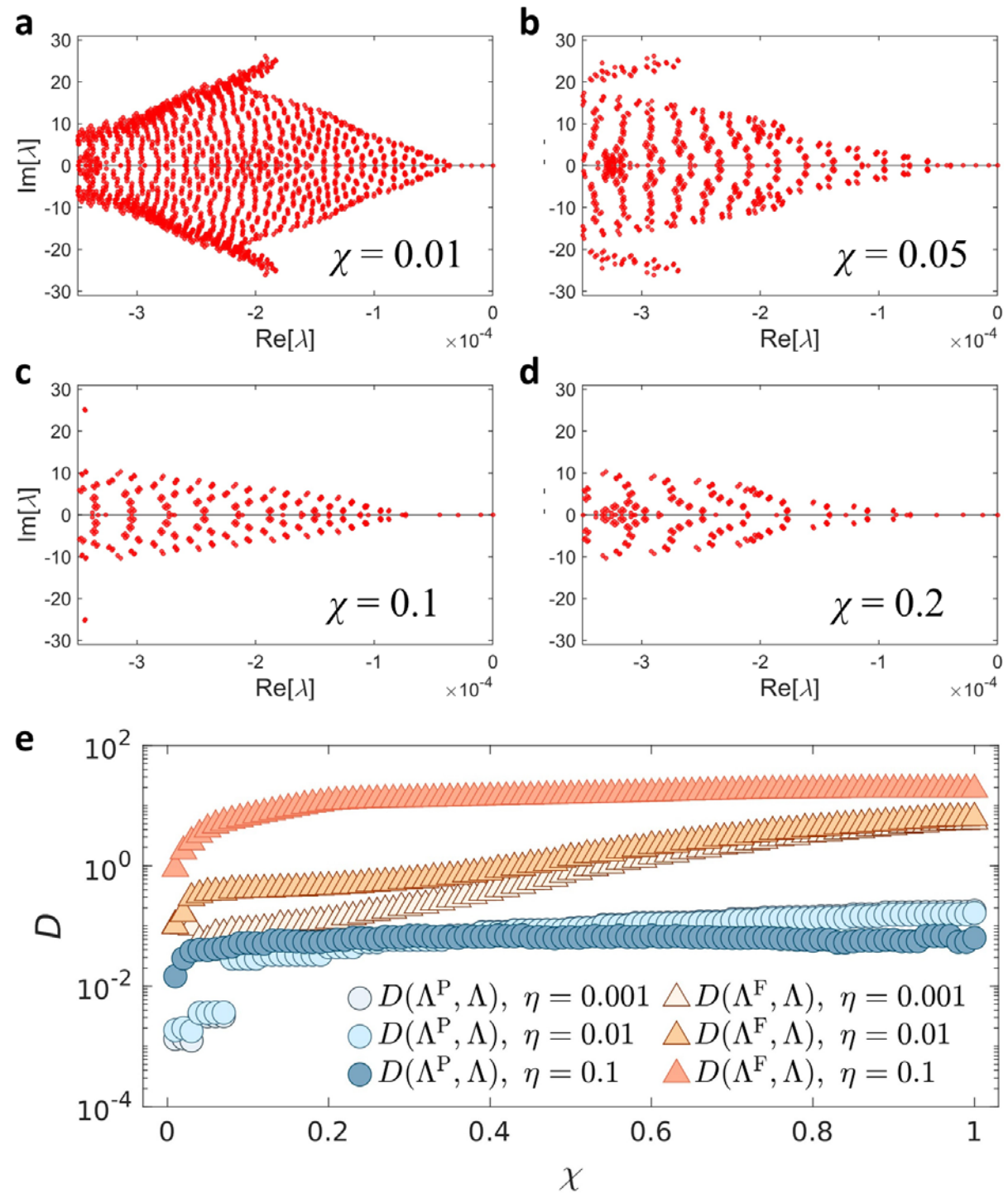

**Supplementary Figure S2. Extended analysis of graph pruning**. Liouville spectra from matrix pruning for $\chi$ = 0.01 (**a**), 0.05 (**b**), 0.1 (**c**), and 0.2 (**d**) at $\eta$ = 0.1. **e**, Chamfer distance $D$ versus $\chi$ for graph pruning (circles) and matrix pruning (triangles) at $\eta$ = 0.001, 0.01, and 0.1. All other parameters are the same as those in Fig. 3 in the main text.

**Supplementary Note S3. Details of GCN**

We generate training data for the open QRM by independently sampling the dissipation rates ($\kappa$, $\gamma$, $\gamma_{\varphi}$) from a log-uniform distribution over [$10^{-6}$, $10^{-4}$], with the Fock-space cutoff set to $N_{\mathrm{max}} = 10$ and the coupling strength $\eta = 10^{-3}$. The resulting $10^4$ samples are split into training, validation, and test sets in a ratio of 8:1:1.

The inputs to the GCN are two magnetic graphs $G_{\mathrm{H}}$ and $G_{\mathrm{L}}$, each of which carries its vertex potential as a node feature (Fig. S3). Because $G_{\mathrm{H}}$ has $N_{\mathrm{c}} = 2(N_{\mathrm{max}} + 1) = 22$ vertices, its subnetwork uses two graph-convolutional layers with a hidden width of 128. In contrast, $G_{\mathrm{L}}$ has $N_{\mathrm{c}}^2 - 1 = 483$ vertices, corresponding to the dimension of the Kossakowski matrix in the traceless operator basis, and its subnetwork therefore employs four graph-convolutional layers with a hidden width of 512. Because both graphs are magnetic, with conjugate-symmetric phase factors assigned to their edges, each layer maintains separate real and imaginary weight matrices.

Among the entire ($N_{\mathrm{c}}^2 = 484$) eigenvalues, we predict the 483 nontrivial eigenvalues, excluding the trivial steady-state mode at $\lambda = 0$. After the convolutional layers, the real and imaginary node features from each subnetwork are aggregated using mean and max pooling, yielding a 2048-dimensional representation for $G_{\mathrm{L}}$ and a 512-dimensional representation for $G_{\mathrm{H}}$. These representations are concatenated into a single 2560-dimensional feature vector, which is passed through a multilayer-perceptron head with hidden widths of 64 and 32. The final output layer has size 966 and is reshaped into the predicted real and imaginary parts of the 483 target complex eigenvalues.

We train a separate GCN with the same architecture for each pruning threshold $\chi = 0$ to 1 in steps of 0.05 (Fig. S3). At $\chi = 0$, the pristine graph $G_{\mathrm{L}}$ is used as input; for $\chi > 0$, the corresponding edges of $G_{\mathrm{L}}$ are pruned to obtain $G_{\mathrm{L}}{}^{\chi}$, following the same pruning procedure as in the main text. In every case, the network is trained to predict the eigenvalues of the original Liouvillian from the input graphs, so that performance across different values of $\chi$ can be fairly compared using a common target. All networks are trained with the Adam optimizer using a batch size of 128 for up to 200 epochs, with an exponentially decaying learning rate from $10^{-3}$ to $10^{-5}$. Because the predicted and true eigenspectra form unordered sets, the loss for each sample is computed as the mean squared error after establishing a one-to-one correspondence between the predicted and true eigenvalues via Hungarian matching.

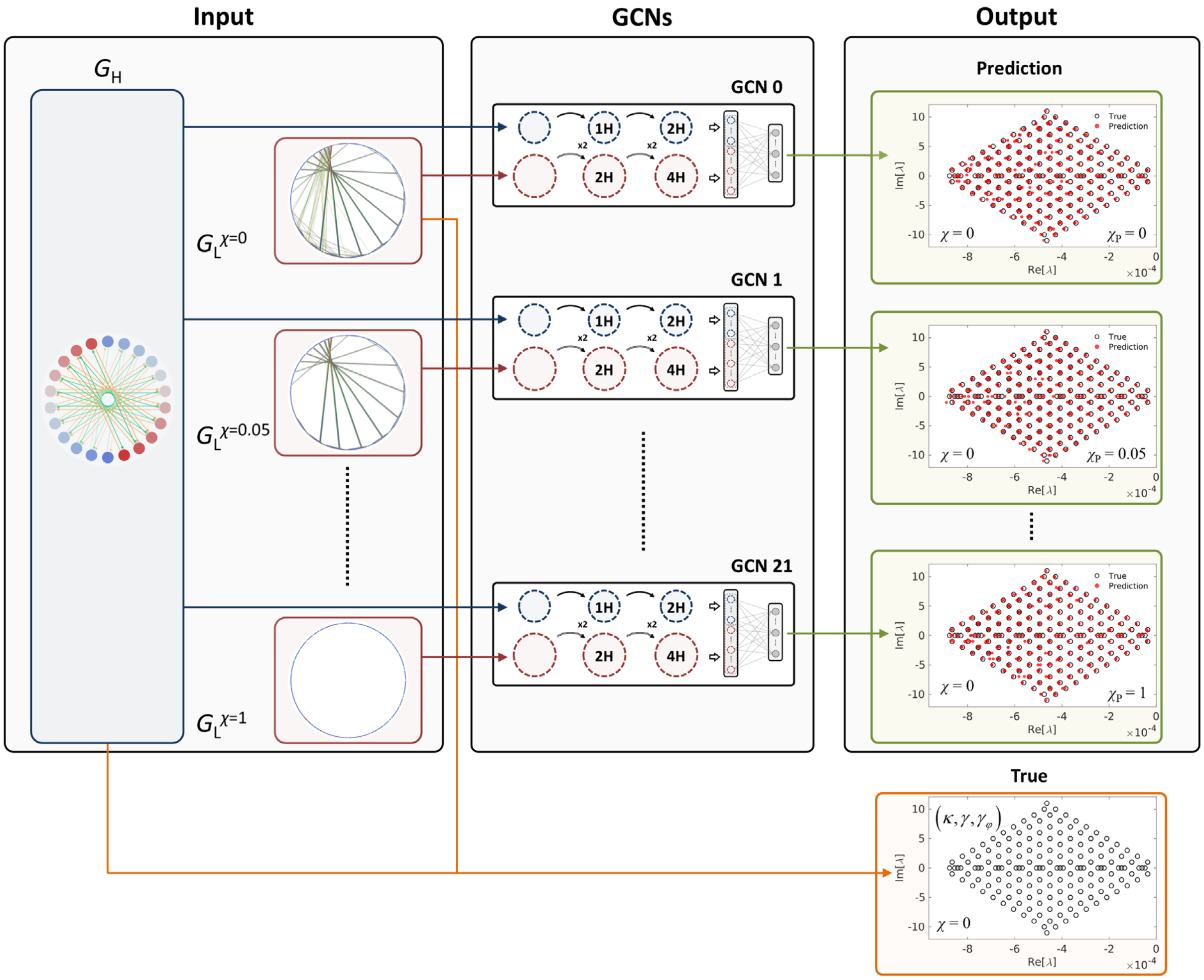

**Supplementary Figure S3. GCN architecture**. Schematic of the GCN architecture used to infer the Liouville spectrum from the input graphs $G_{\mathrm{H}}$ and $G_{\mathrm{L}}{}^{\chi}$, with a separate GCN trained for each pruning threshold $\chi$ against the common target spectrum of the pristine ($\chi = 0$) Liouvillian spectrum. Representative predicted spectra (red dots) are compared with the true eigenvalues (open circles) for increasing $\chi$.

**Supplementary Note S4. Examples of GCN prediction**

For each pruning threshold $\chi$, we select the test samples with the smallest and largest Chamfer distances between the true and predicted Liouville spectra and overlay the two spectra in Figs. S4 and S5, respectively. Across the full range of $\chi$, the predicted spectra in the best scenarios in Fig. S4 closely reproduce the diamond-shaped distributions of the corresponding true spectra.

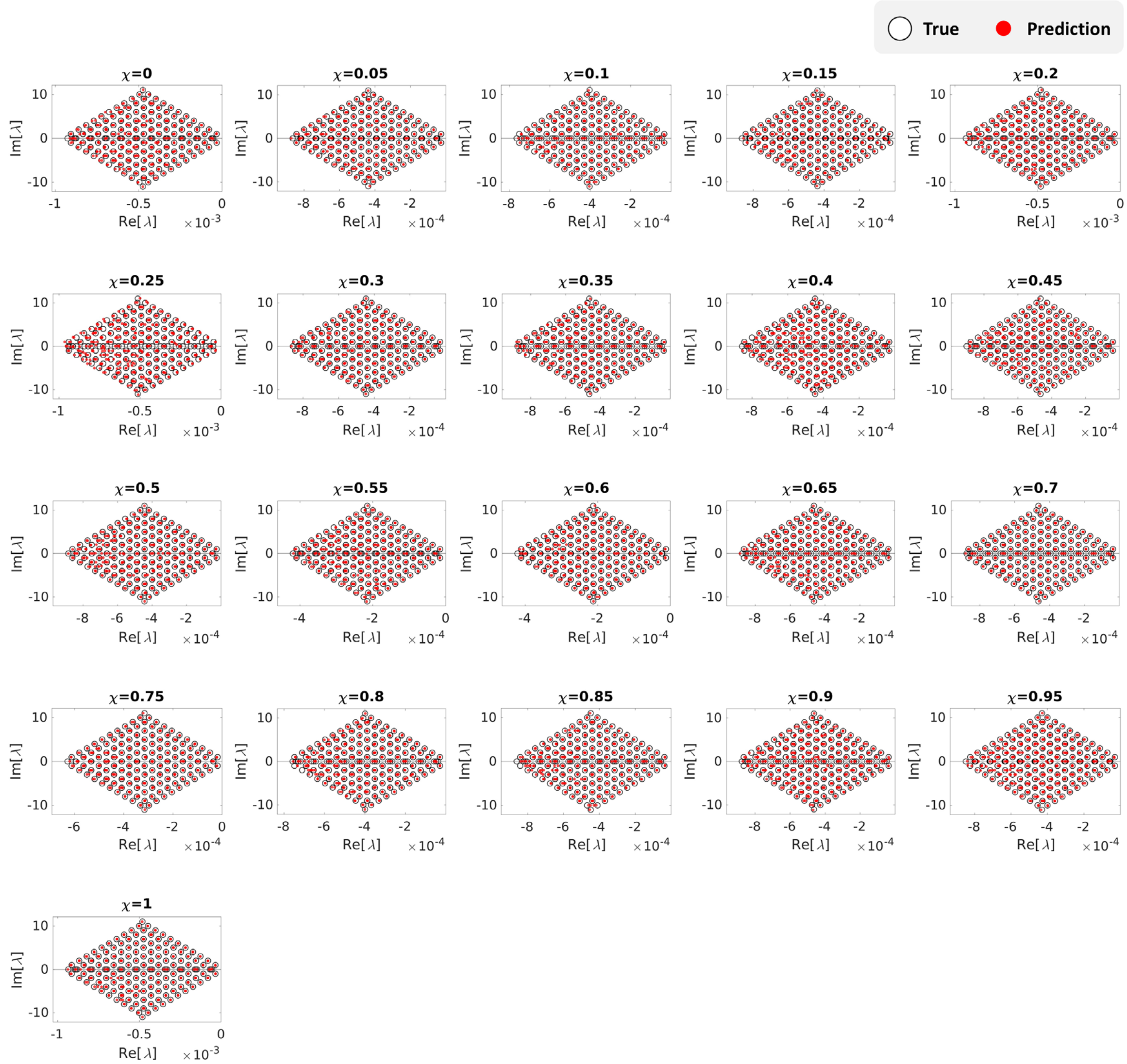

**Supplementary Figure S4. Best-case GCN predictions.** Best-case true (open circles) and predicted (filled circles) Liouville spectra, selected by the smallest Chamfer distance $D$, for pruning thresholds $\chi = 0$ to 1 in steps of 0.05. The corresponding $\chi$ value is labelled in each panel.

In contrast, the predicted spectra for the worst-case scenarios in Fig. S5 show distinct discrepancies from the true spectra depending on the value of $\chi$: a spectral shift at lower $\chi$ values and shape deformation at higher $\chi$ values. This distinction mainly originates from the general tendency toward spectral broadening at higher $\chi$, which corresponds to the effect of reduced dimensionality on the universal spectral responses of Markovian open quantum systems[1]. Therefore, our pruning method functions by removing less significant dimensions in the quantum master equation.

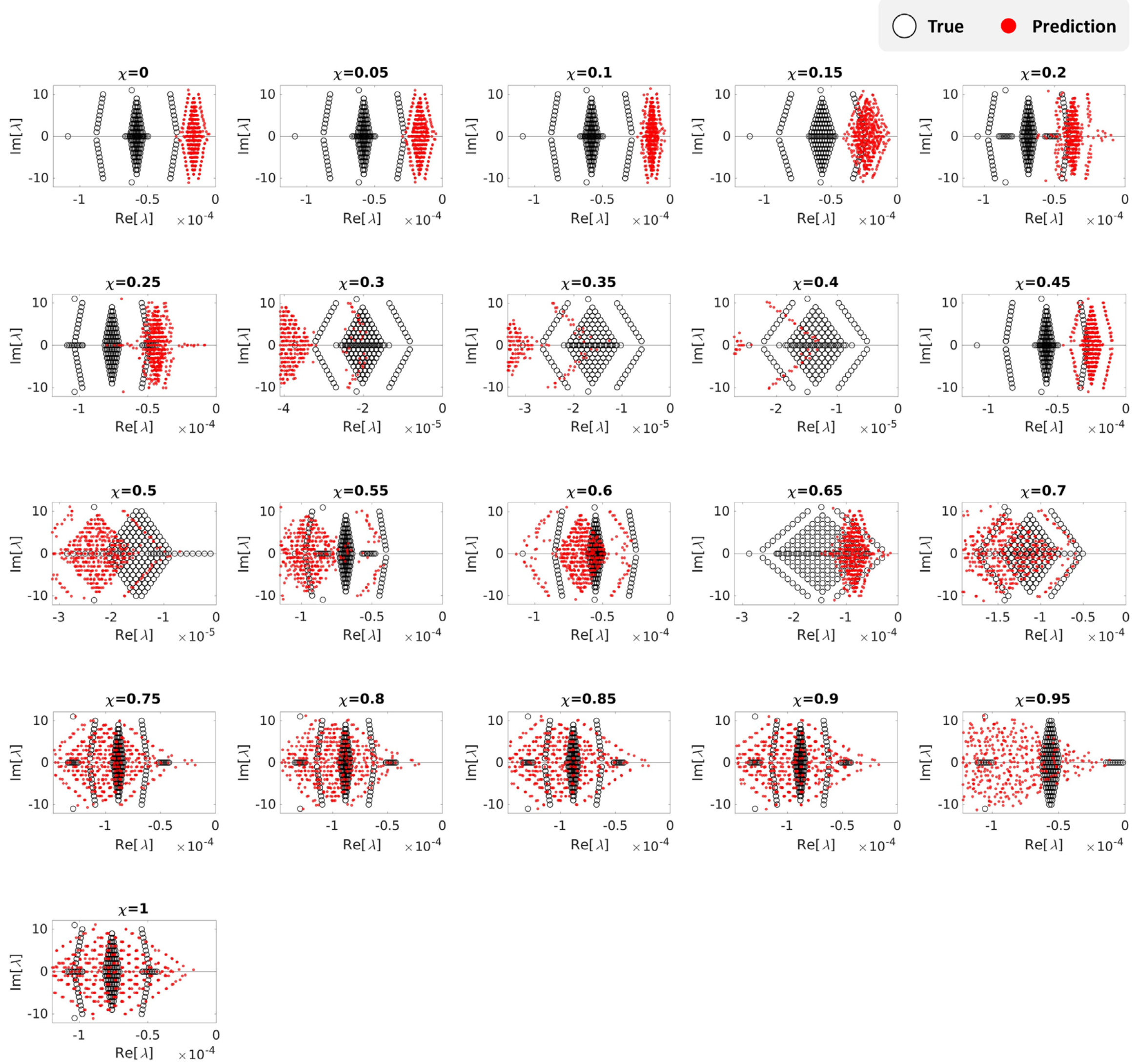

**Supplementary Figure S5. Worst-case GCN predictions.** Worst-case true (open circles) and predicted (filled circles) Liouville spectra, selected by the largest Chamfer distance $D$, for pruning thresholds $\chi = 0$ to 1 in steps of 0.05. The corresponding $\chi$ value is labelled in each panel.

**Supplementary Note S5. Training curves**

Figure S6 shows the resulting training and validation loss curves for each pruning threshold $\chi$. For all cases, the training and validation losses converge closely and decrease monotonically, indicating stable training without appreciable overfitting. Notably, pruning with $\chi \geq 0.15$ allows the model to escape the learning plateau near the loss value $L = 3 \times 10^{-2}$. This suggests that weak weighted edges partly act as noisy connections in the learning of graph convolution. By removing such edges, pruning reduces spurious message passing and improves the effective signal-to-noise ratio of the graph representation, thereby enabling more efficient learning of the Liouville spectrum.

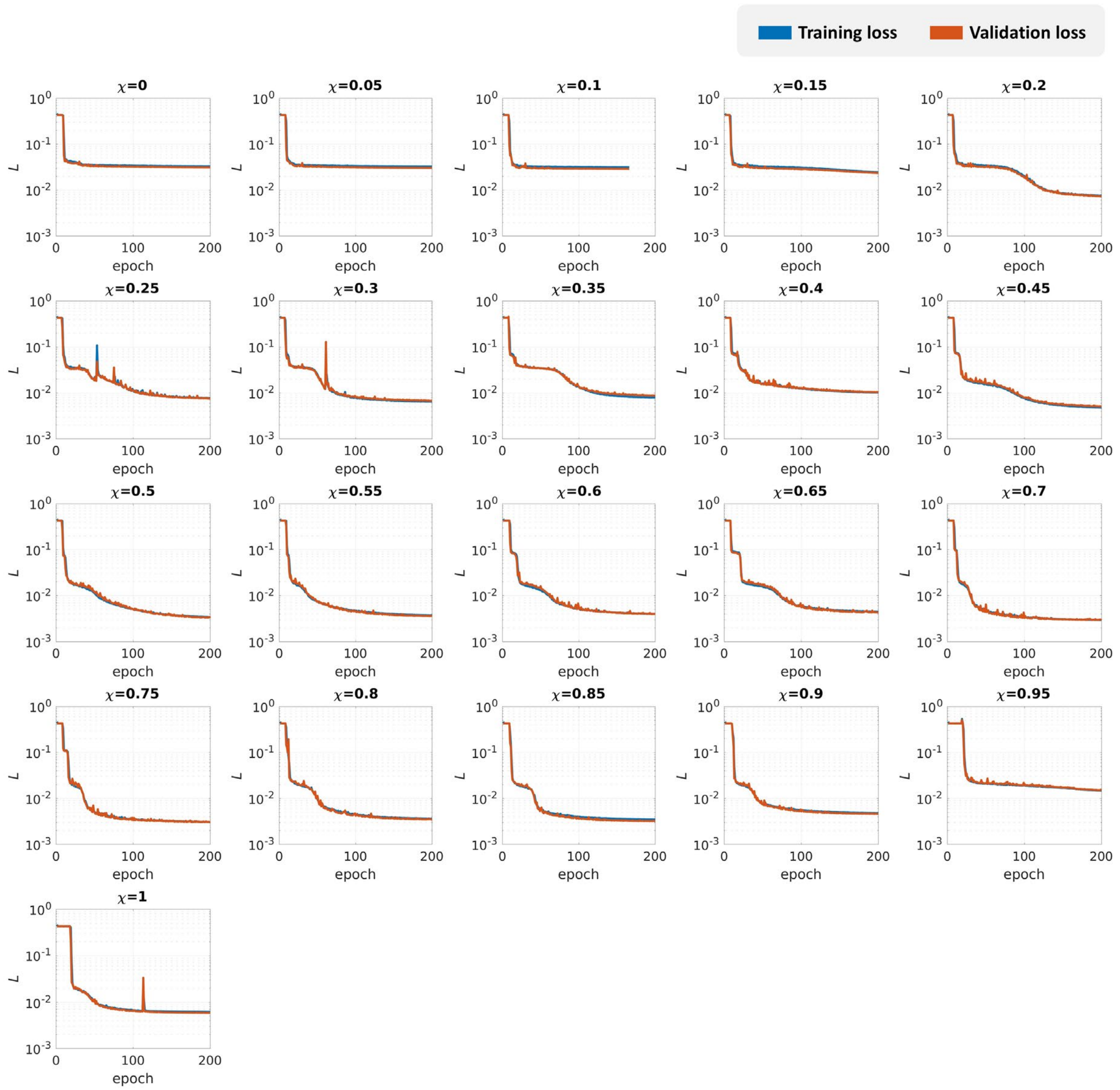

**Supplementary Figure S6. Training and validation loss curves.** Training (blue) and validation (orange) losses versus epoch for GCNs trained on $G_L^{\chi}$ with pruning thresholds $\chi = 0$ to 1 in steps of 0.05.

## Supplementary References